\documentclass[conference]{IEEEtran}
\IEEEoverridecommandlockouts
\usepackage{cite}
\usepackage{amsmath,amssymb,amsfonts,amsthm}
\usepackage{graphicx}
\usepackage{textcomp}
\usepackage{xcolor}

\usepackage[utf8]{inputenc}
\usepackage{booktabs}
\usepackage{algorithm}
\usepackage{booktabs}
\usepackage{multicol}
\usepackage{algpseudocode}
\usepackage{amssymb}
\usepackage{subfig}
\usepackage{todonotes}
\usepackage{multirow}
\usepackage{bm}
\usepackage[switch]{lineno}
\usepackage{array}
\usepackage{tabularx}
\usepackage{tabulary}
\usepackage{makecell}

\makeatletter
\def\ps@IEEEtitlepagestyle{%
  \def\@oddfoot{\mycopyrightnotice}%
  \def\@evenfoot{}%
}
\def\mycopyrightnotice{%
  {\footnotesize 978-1-6654-3156-9/21/\$31.00
©2021 IEEE \hfill}
  \gdef\mycopyrightnotice{}
}

\newenvironment{conditions*}
  {\par\vspace{\abovedisplayskip}\noindent
   \tabularx{\columnwidth}{>{$}l<{$} @{${}={}$} >{\raggedright\arraybackslash}X}}
  {\endtabularx\par\vspace{\belowdisplayskip}}

\def\BibTeX{{\rm B\kern-.05em{\sc i\kern-.025em b}\kern-.08em
    T\kern-.1667em\lower.7ex\hbox{E}\kern-.125emX}}
    
\begin{document}

\title{Blind Calibration of Air Quality Wireless Sensor Networks Using Deep Neural Networks}

\author{\IEEEauthorblockN{Tiago Veiga\IEEEauthorrefmark{1}, Erling Ljunggren\IEEEauthorrefmark{1}, Kerstin Bach\IEEEauthorrefmark{1}, Sigmund Akselsen\IEEEauthorrefmark{2}}
\thanks{This work was funded by the European Union’s Horizon 2020 research and innovation program, project AI4EU, grant agreement No. 825619.}
\IEEEauthorblockA{\IEEEauthorrefmark{1} Norwegian University of Science and Technology, Dept of Computer Science\\
Trondheim, Norway\\
Email: \{tiago.veiga,kerstin.bach\}@ntnu.no, erlingljunggren@gmail.com}
\IEEEauthorblockA{\IEEEauthorrefmark{2}Telenor Research \\ Oslo, Norway \\
Email: sigmund.akselsen@telenor.com}
}



\maketitle

\IEEEpubidadjcol

\begin{abstract}
Temporal drift of low-cost sensors is crucial for the applicability of wireless sensor networks (WSN) to measure highly local phenomenon such as air quality. The emergence of wireless sensor networks in locations without available reference data makes calibrating such networks without the aid of true values a key area of research. While deep learning (DL) has proved successful on numerous other tasks, it is under-researched in the context of blind WSN calibration, particularly in scenarios with networks that mix static and mobile sensors. In this paper we investigate the use of DL architectures for such scenarios, including the effects of weather in both drifting and sensor measurement.
New models are proposed and compared against a baseline, based on a previous proposed model and extended to include mobile sensors and weather data.
Also, a procedure for generating simulated air quality data is presented, including the emission, dispersion and measurement of the two most common particulate matter pollutants: PM\textsubscript{2.5} and PM\textsubscript{10}.
Results show that our models reduce the calibration error with an order of magnitude compared to the baseline, showing that DL is a suitable method for WSN calibration and that these networks can be remotely calibrated with minimal cost for the deployer.
\end{abstract}

\begin{IEEEkeywords}
Sensor networks, Sensor blind calibration, Deep learning, Air quality
\end{IEEEkeywords}

\section{Introduction}
Air pollution poses a major threat to health and climate with a high economic impact in several countries.
A solution to accurate air quality monitoring is the deployment of wireless sensor networks (WSN) \cite{Kumar2015,veiga2020}. In the context of this paper, a WSN is a set of low-cost sensors that enables large-scale local measuring, as they are cheap enough to be placed densely over a large area. Moreover, in urban areas, air quality is a local phenomen making it difficult to monitor with fixed sensors. Mobile sensors (e.g., mounted on city buses) provide additional data to improve measures and predictions of air quality. Even though cheap WSNs improve the data acquisition process, its maintenance, accuracy, and reliability remain a challenge. Therefore, maintenance costs are a concern, in particular regarding the need for service, calibration and the limited lifetime of such devices. We present solutions addressing the automatic calibration of air quality sensors as a mean to reduce the maintenance costs of such a WSN. 

The calibration of sensors increases data quality, which might suffer due to various causes \cite{Fang2017}, but most importantly by accumulating larger, varying, drift rates as they age. Frequent calibration of sensors is problematic for WSNs due to the sheer amount of sensors usually deployed, leading to lot of work to either ship sensors back to the lab or carry equipment for local calibration on its mounting location.
An alternative is to calibrate such sensor networks remotely. For calibrating WSNs remotely, one important factor is how many high-quality reference-nodes are available. Ideally, the calibration procedure should be feasible without any, or at least requiring only a few. This is called blind or partially blind calibration, and is becoming an important research topic since it allows high quality measurements with less expensive sensors.

To the best of our knowledge, deep learning (DL) for blind WSN calibration is an under-researched field, with one exception which only considers a network of static sensors \cite{Wang2017}.
It proved better than non-DL methods, thus leading to the hypothesis that a good DL model can be created for blind calibration. We leverage from previous work, extending DL models for scenarios with static and mobile sensors and presenting three newly designed models, with designs based off of key advances in related fields of DL tailored to the calibration problem. The three models use convolutions in one dimension, convolutions in two dimensions, and LSTMs with attention as their key components.


Originally, this work's goal was to use data from an air quality monitoring sensor network to be deployed in an urban scenario. However, due to the impact of the COVID-19 pandemic the deployment suffered delays and, despite starting to become available, there wasn't enough data in order to properly train models. Moreover, to the best of our knowledge, there is no available public datasets that combine static and mobile sensors. Therefore, we also propose a synthetic data generation method to evaluate the developed architectures.

The main contributions of this paper are: a method to generate simulated pollution data for static and mobile sensors, including the influence of meteorological data; a deep learning approach for blind calibration of static and mobile sensors.

\section{Related Work}
WSNs have seen great success in monitoring air quality on a local level \cite{Maag2018}, with increased monitoring accuracy with a fine-grained prediction model using WSNs as data source instead of a few accurate sensors \cite{Kumar2015}.
Still, the accuracy of such networks depends on proper sensor calibration, preferably without removing them from the deployment location \cite{Delaine2019}.

Two important methods are subspace projections and strategies based on consensus. Projection methods map the sensor data to a sub-dimensional space where the drift is recoverable by leveraging assumptions on the nature of the drift phenomenon and measurand. Consensus algorithms are mostly used on mobile sensors where rendezvous can be used to update some pre-defined calibration parameters by comparing the sensor outputs. This method was used on static dense networks by extending assumptions on how the measurand disperse \cite{Stankovic2018}. Other methods include the use of Bayesian models by \cite{YangA2018} and \cite{YangJ2018}, that model the phenomenon as a known Gaussian process and leveraging assumptions on the drift. \cite{Becnel2019} uses a recursive definition on the calibration relationships between the sensors in a network, propagating the attributes of reference sensors in the network to calibrate other sensors.

To the best of our knowledge, only one author \cite{Wang2017} designed DL methods for blind calibration, with a convolutional network that mapped the sensor data into a subspace defined by convolutional kernels, similar to other subspace projection methods, and then retrieved the drift-free measurements using stacked convolutions. Using DL relied less on assumptions, because of the flexibility of the model, and performed better than compared models. Later \cite{YangA2018} outperformed the model using Gaussian processes and explicit long-term dependencies. However, none of them experimented the extra challenge of calibrating mobile sensors.


\section{Problem Definition}

\subsection{WSN drift and calibration}

Calibration is defined as deriving the relationship between the raw output of the sensors and the real quantity measured by the sensors.
Consider n sensors, labeled $i = 1, 2, ..., n$, measuring a continuous signal at $T$ discrete timestamps labeled $t = 1, 2, ..., T$. A sensor output $y_{i,t}$ and its corresponding real measurement $x_{i,t}$ are then correlated as defined by the following equation.

\begin{equation}
    \label{eq:sensor_error_full}
    y_{i,t} = \alpha_{i,t} x_{i,t}^{\beta_{i,t}} + c_{i,t} + \epsilon_{i,t}
\end{equation}

Where $\alpha$ is the linear part of the error, $\beta$ is the non-linear part, and $c$ is the constant part. $\epsilon$ defines the random noise at each measurement. Note that drift variables are dependant on time because the drift variables are dependant on the age of the sensor, but also history and exogenous variables, with history being the previous sensor readings and exogenous variables all the external variables that influence the readings (for instance, meteorological data). Given the complexity of the problem, simplification is made by noting that manufacturers tend to correct for non-linearity with on-chip post-processing, \eqref{eq:sensor_error_full} can be simplified by removing $\beta$ as follows:

\begin{equation}
    \label{eq:sensor_error_simple}
    y_{i,t} = \alpha_{i,t} x_{i,t} + c_{i,t} + \epsilon_{i,t}
\end{equation}

Which is the equation most used in the literature. Further simplifications can be made
by assuming the exogenous variables have little effect on the measurement error
and ignore them, and by ignoring the effect of aging and temporal differences. This results in four main schools of calibration, utilizing relationships of varying complexity, by employing none, either one or both of the mentioned simplifying assumptions.
The goal of calibration is then to find a function f (·) that minimizes the difference
between all measured and real values.

\begin{equation}
    \label{eq:calibration_optimization} 
    \min_{f(\cdot)} \sum_i \sum_t |f(\mathbf{y_{i,t}}) - \mathbf{x_{i,t}}|
\end{equation}

Where $|\cdot|$ denotes absolute value and all $x_{i,t}$ are unknown in the case of blind calibration. For partially blind problem specifications some sensors are known to be correct.

\subsection{Calibration as a time series problem}
Calibration can be viewed as a sequence-to-sequence (Seq2Seq) problem, where a time series (TS) is mapped to another.
While Seq2Seq problems often correlate two TS on a series level, e.g. translation, here there is a direct mapping between each datapoint from input and output.

Similarities are also shared with time series forecasting (TSF), related to the causality of the TS and the target data. The causality of a TS (i.e., the defined ordering of the datapoints) is important for TSF as we predict for time $t+1$ using $t$, and for calibration as we are interested in the newest datapoint $t$. While using future values could improve performance, it delays the best predictions.
For the target data the same reasoning applies, as it closely relates to each datapoint, being it either the following one (for TSF) or the calibrated version (Calibration).

As a summary, calibration maps between two TS of identical length with a direct mapping on a datapoint level, where causality is important. Models leveraging the direct mapping between datapoints or timesteps should be considered.

\section{Model Architectures}

\begin{figure*}[tb]
    \centering
     \subfloat[Baseline: adapted from \cite{Wang2017}.]{\includegraphics[width=0.5\textwidth]{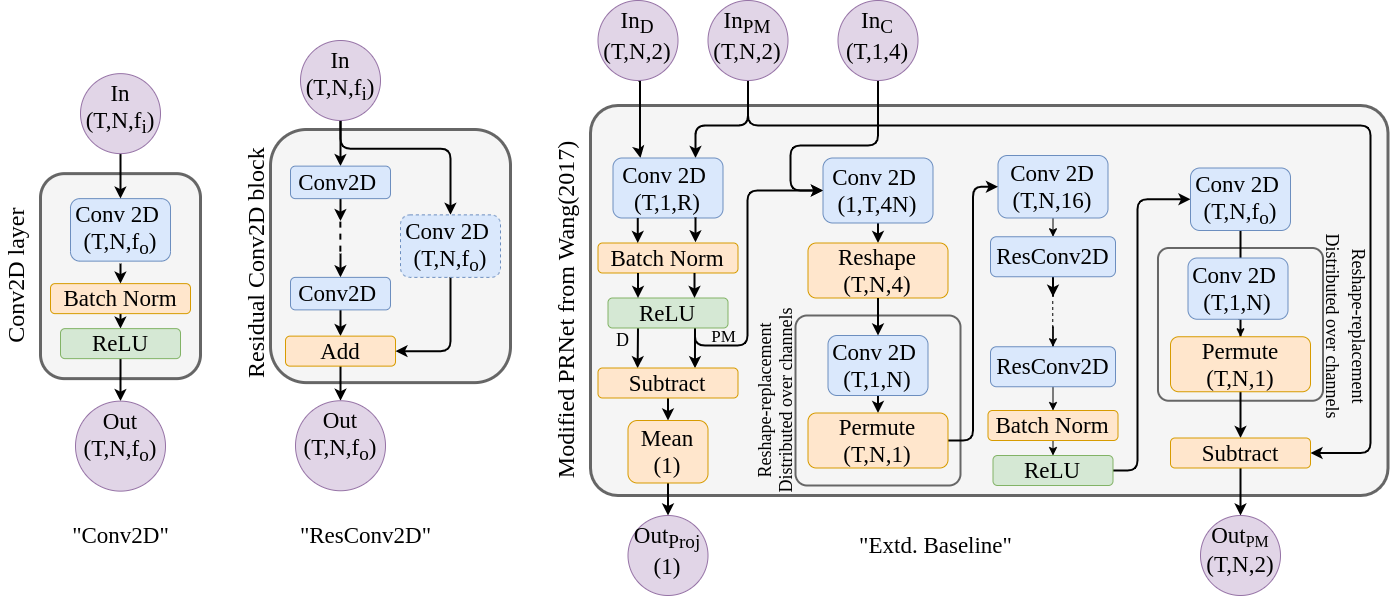}\label{fig:baseline_archt}} \hspace{15pt}
     \subfloat[Conv1D: the convolutional model using 1D convolutions.]{\includegraphics[width=0.4\textwidth]{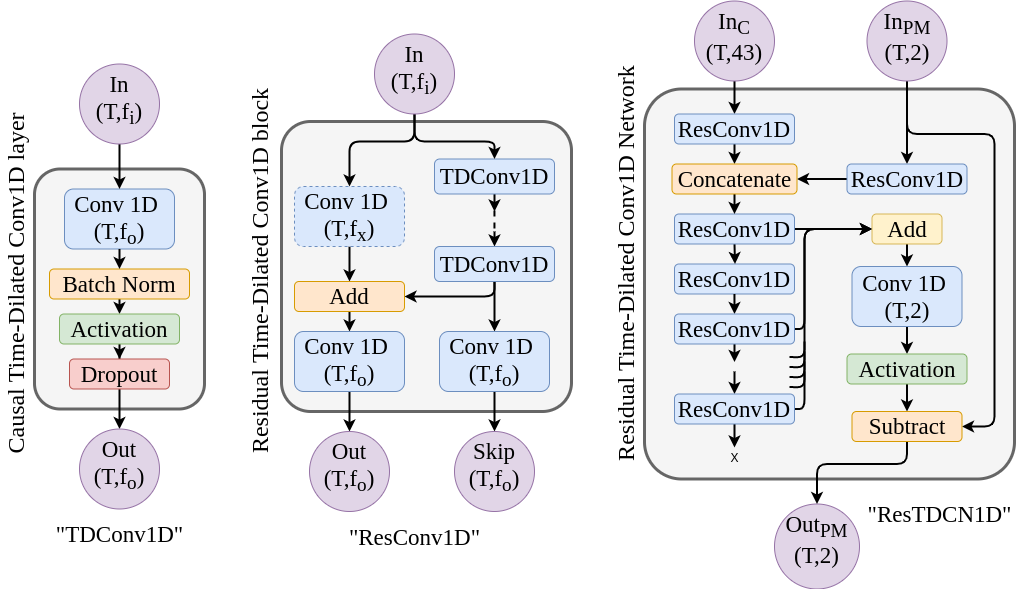}\label{fig:conv1d_archt}} \\
     \subfloat[Conv2D: the convolutional model using 2D convolutions.]{\includegraphics[width=0.45\textwidth]{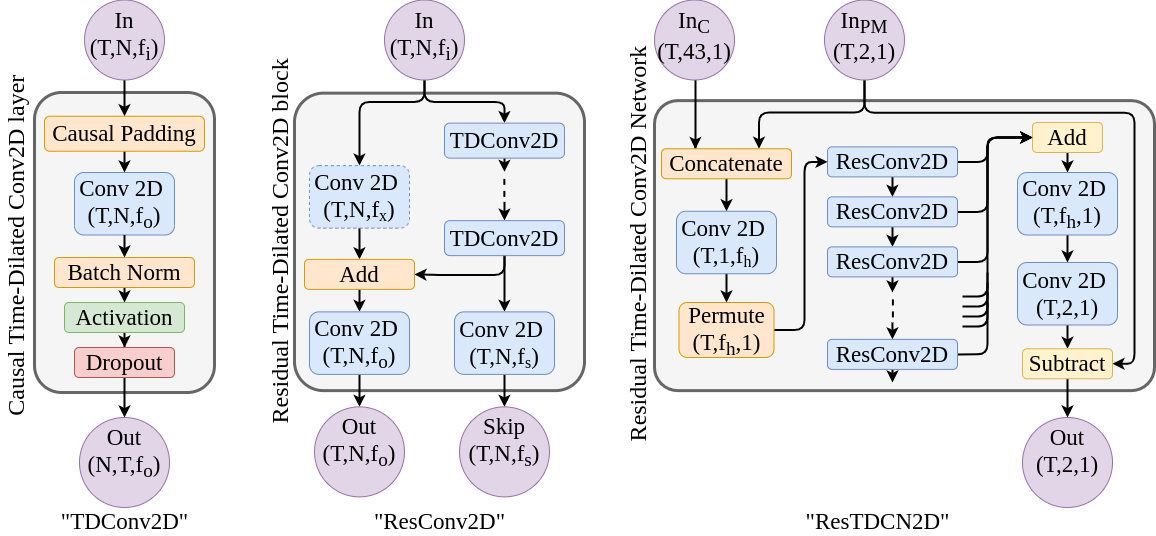}\label{fig:conv2d_archt}} \hspace{15pt}
     \subfloat[LSTMwA: a stacked LSTM with a convolutional attention mechanism.]{\includegraphics[width=0.5\textwidth]{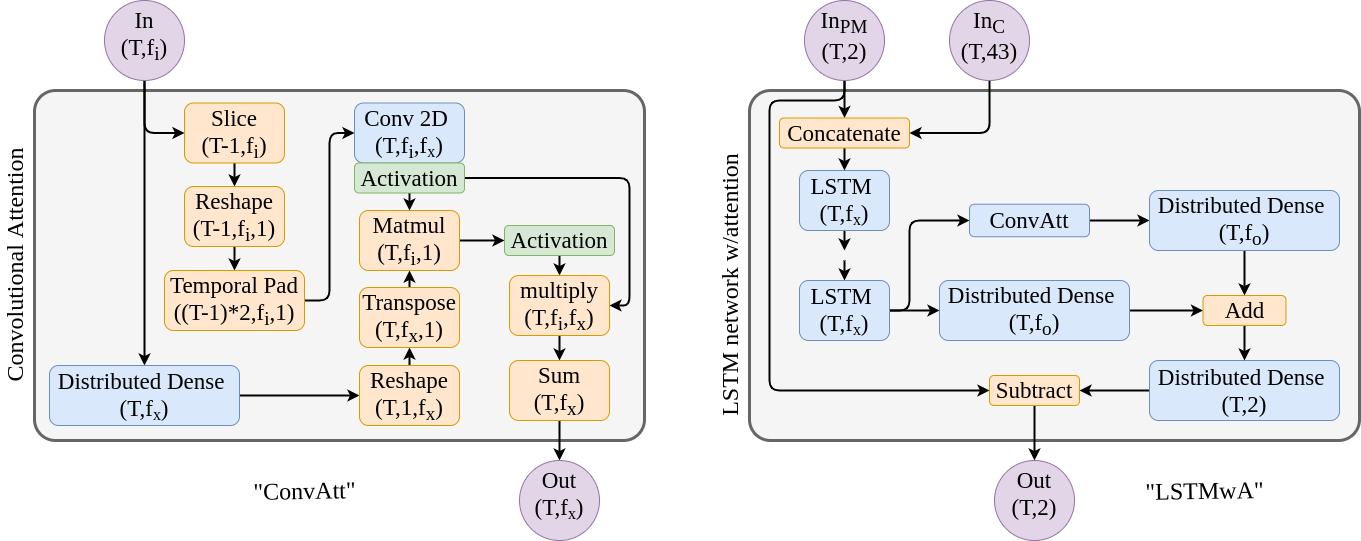}\label{fig:lstm_archt}}
    \caption{Model architectures used in WSN calibration. Inputs and outputs are denoted by subscript ''C'' for context input, ''PM'' for PM input,  ''D''for drift, input nodes without subscript are general. Output shapes are given for layers changing them if they are not defined in the figure. All filter dimensions, shown as f, are tuneable, and T and N are extrapolated from data.}
    \label{fig:model_archt}
\end{figure*}

\subsection{Baseline}

The baseline model derives from the only DL model originally designed for blind WSN calibration \cite{Wang2017}, which 
implements a projection as a convolution with a kernel spanning the non-temporal dimensions, followed by a ResNet architecture. Simultaneously PM\textsubscript{10} and PM\textsubscript{2.5} are calibrated by adding the values as an extra channel in the input matrix. Exogenous values are concatenated with the projection output in the channel dimension before the ResNet layers.
The implementation of the "Rearrangement" layer is implemented as a convolution similar to the projection layer to facilitate mobile sensors.

\subsection{Conv1D}

The baseline has a short temporal "memory", and we designed a calibration network based on the WaveNet architecture \cite{wavenet} to alleviate this, as the temporal dimensions enable modelling longer temporal dependencies. Before applying the temporally dilated 1D-convolutions, we process the exogenous variables and the sensor output separately with a non-dilated convolutional layer. The network output is subtracted from the sensor measurements to produce predictions for the true measurements, as this focuses computational power on drift modelling. 
The output of the WaveNet architecture is reduced using a convolutional layer on the sum of residual skip-connections with a kernel size
that is greater than the kernels used in the residual blocks, with a daily dilation rate. 

\subsection{Conv2D}

In order to extend the analysis of dilated convolutions, we extended the previous model with an extra dimension, with small changes to support this extra dimension better. The initial residual units have been replaced by a convolutional layer spanning the entire variable space, similar to the projection convolution of the baseline model. There are also two distinct convolutional layers to reduce the residual output, one for filters and one for variables, with the variable-reducing convolution utilizing a larger kernel with a daily dilation rate.

The temporally causal 2D convolutions were obtained by selecting the first dimension as temporal and only dilating in that dimension. Causality is obtained by causal padding in that dimension, only for past variables, while padding normally in both directions for the variable dimension. 
This keeps the input shape, enabling use of residual connections.

\subsection{LSTM w/ Attention}

Recurrent models are another important model for TS tasks, and we implemented a stacked LSTM with attention to complete our initial exploration of DL models for calibration. Similarly to the other models, we subtract the network output from the raw sensor measurements.

The attention layer is a convolutional attention designed by \cite{Shih2019} that attend convolutional filter-values from all past timesteps, modified to generate output for all timesteps. The context vectors are obtained by a causal 2D convolution on all timesteps up until the next newest one, as designed for the Conv2D network, with a kernel spanning the temporal dimension. The query vectors are obtained from a distributed dense layer with number of nodes equal the number of convolutional filters.
Due to 
explosive gradients, we clip the gradients to a norm
of 0.9 for all experiments with this model.

\section{Data Generation}

\begin{algorithm}[tb]
  \caption{Synthetic data generation}
  \label{alg:synt_data}
  \begin{algorithmic}[1]
  \State Place pollution sources and static sensors
  \State Compute path of mobile sensors
  \ForAll{t in timesteps}
    \State Sample weather
    \ForAll{s in sources}
        \State Compute emissions $x^s_t$ (Eq.\ref{eq:source_emissions})
    \EndFor
    \ForAll{z in sensors}
        \State Compute drifted measurement $y^z_t$ (Eq.\ref{eq:sensor_measurements} and Eq.\ref{eq:synth_drift})
    \EndFor
  \EndFor
  \end{algorithmic}
\end{algorithm}

We developed a procedure to generate simulated air quality data. To the best of our knowledge, no public datasets were available at the necessary sensor density, with mobile sensors and meteorological data. Also, in the real world it can be difficult to find accurate ground truth measurements, as the most accurate sensors are expensive and hard to maintain so only a few of them might be available to serve as references.

We simulate emission, dispersion and measurement of the two most common measured particulate matter (PM) sizes: PM\textsubscript{2.5} and PM\textsubscript{10}.
Algorithm~\ref{alg:synt_data} resumes this procedure. It computes random locations for sources and static sensors, and paths for mobile sensors. Then, for each timestep, it samples weather at each location and computes the emissions and sensor readings, with the respective influence of meteorological variables and each emission source.

\subsection{Source and Sensor location}

Location of sources is randomly placed within a circle with radius 100 based on a uniform distribution with a minimum distance of 12 between them. Static sensors follow the same procedure with radius 80. The paths of mobile sensors are sampled as 5 to 15 random points on a circle with varying radius for each location in the path, the center of which is sampled with the same procedure as static sensors.

\subsection{Source Emissions}

Hourly PM values were sampled as a random walk according to the following equations in three steps: the initial sampling $\dot{x}$, the exploded series $\hat{x}$, and the final values $x$. Raising $\dot{X}$ to the power of $7$ is done to introduce pollution spikes, making scaling the values down to $[0,50]$ afterwards necessary. The final sampled TS $X$ is added to sine waves with weekly, monthly and yearly frequencies.

\begin{subequations}
    \label{eq:source_emissions}
    \begin{align}
        \dot{x}_{t+1} & = \text{abs}(\dot{x}_t + \delta_{t}) \\
        \hat{X}_{[\tau_i, \tau_{i+1}\rangle} & = \dot{X}_{[\tau_i, \tau_{i+1}\rangle}^{7} \\
        X_{[\tau_i, \tau_{i+1}\rangle} & = \frac{m_i}{\max \hat{X}_{[\tau_i, \tau_{i+1}\rangle} } \cdot \hat{X}_{[\tau_i, \tau_{i+1}\rangle} \\
        \delta_t & \sim \mathcal{N}(\mu = -0.01\dot{x}_t, \sigma = 1) \in [-1, 10] \\
        m & \sim \mathcal{N}(\mu = 50, \sigma = 9) \\
        \forall_i & \;\,\tau_{i+1} - \tau_i = 30 * 24
    \end{align}
\end{subequations}

\subsection{Weather}
Temperature and humidity were sampled by a random walk centered around $10$ and $80$ respectively. Wind speed was sampled similar to source emissions with an exponential factor of $2$. Wind direction was a modular random walk with uniform steps between $[-60,60]$.

\subsection{Sensor measurements}

Sensor measurements are computed by adding the individual measurements from each source multiplied with a distance coefficient similar to the approach of \cite{Wang2017}. Additionally, we introduce meteorological influence in pollution spreading and, therefore, the distance coefficient has been changed to factor in how the wind blows, in addition to using past PM values for sources distant from the sensor. The true output $y$ for a given sensor $i$ at time $t$ ($y_{i,t}$) is decided by the following equations.

\begin{subequations}
    \label{eq:sensor_measurements}
    \begin{align}
        o_{i,c,t} =& \lfloor \frac{d_{i,c,t}}{\frac{2}{5}R} \rfloor  \\
        w_{c,i,t} =& \frac{s_{t}}{o_{i,c,t}} \sum_{\tau = 0}^{o_{i,c,t}} 2 (1-\frac{(\phi_{(c,i),t-\tau}-\phi_{w,t-\tau})\mod \pi}{\pi}) - 1 \\
        a_{i,c,t} =& \begin{cases}
            (\frac{10 d_{i,c,t}}{\frac{R}{2} + \frac{R}{2}(w_{i,c,t}+1+(2^{\frac{1}{2}} w_{i,c,t})^2)} +1)^{-1.5} & \text{if $w_{i,c,t} > 0 $} \\
            (\frac{10 d_{i,c,t}}{\frac{R}{2} + \frac{R}{2}(w_{i,c,t}+1)} +1)^{-1.5} & \text{otherwise} \\
        \end{cases} \label{eq:measuring_coefficient}\\
        y_{i,c,t} =& a_{i,c,t} e_{c,t-o_{i,c,t}}\\
        y_{i,t} =& \sum_{c=0}^C y_{i,c,t}
    \end{align}
\end{subequations} 
Where $R$ is the radius of the system, $d$ is the distance between the source and the sensor, $o$ is the offset (which timestep to use when measuring), $\phi_{c,i}$ is the angle between the source and sensor, $\phi_{w}$ is the angle of the wind, $s$ is the wind speed, $w$ is the wind coefficient deciding how the wind affects the measurement, $a$ is the measurement coefficient, $e$ is the emitted PM value, $y_{i,c}$ is the measurement of sensor $i$ from the value for source $c$, and $y_i$ is the total measured PM value.


\subsection{Sensor drift}

We know from the research of \cite{Maag2018} that the errors of the sensors depend on other phenomena, like weather. Therefore, we designed a drift model based on equation \eqref{eq:sensor_error_full} that includes meteorological influence in the sensor drift. The final drifted output $x$ is then defined as follows:

\begin{subequations}
    \label{eq:synth_drift}
    \begin{align}
        \alpha_{i,t} = & f_{\alpha, i} \cdot \textit{$\Tilde{T}$}_{\alpha, i,t} \cdot \textit{$\Tilde{H}$}_{\alpha,i,t} \cdot \textit{$\Tilde{D}$}_{\alpha,i,t} \\
        \beta_{i,t} = & f_{\beta, i} \cdot \textit{$\Tilde{T}$}_{\beta, i,t} \cdot \textit{$\Tilde{H}$}_{\beta,i,t} \cdot \textit{$\Tilde{D}$}_{\beta,i,t}\\
        c_{i,t} = & f_{c, i} + \textit{$\Tilde{T}$}_{c, i,t} + \textit{$\Tilde{H}$}_{c,i,t} + \textit{$\Tilde{D}$}_{c,i,t} \\
        x_{i,t} = & ((1-\tau_{i,t}) + (\tau_{i,t} \alpha_{i,t})) y_{i,t}^{(1-\tau_{i,t}) + (\tau_{i,t} \beta_{i,t})} + \tau_{i,t} c_{i,t} + \epsilon_{t}
    \end{align}
\end{subequations}

Where $\epsilon$ is the random error noise, $c$ is the constant error source, $\alpha$ is the linear error source, and $\beta$ is the exponential error source. $f$ is the independent factor for the error source defined by the subscript, and $\tau$ is the temporal factor deciding how drifted the sensor $i$ is at time $t$, $\Tilde{T}$ is the scaled temperature, $\Tilde{H}$ is the scaled humidity and $\Tilde{D}$ is the scaled history.

The variable $\tau$ increases linearly with the factor $r_{\tau, i}$ sampled for each sensor and clipped such that all values for $\tau$ is smaller than or equal $1$. The scaled values are obtained by scaling the simulated values in random ranges sampled from $[0.95, 1.05]$, $[0.99, 1.01]$, and $[-0.2, 0.2]$ for $\alpha$, $\beta$, and $c$ respectively.

\subsection{Context}

\begin{figure}
    \centering
     \subfloat[Partitioning]{\includegraphics[width=0.27\linewidth]{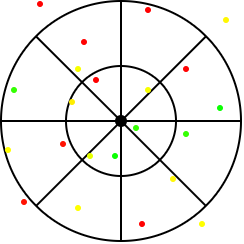}\label{fig:context_partitioning}} \hspace{15pt}
     \subfloat[Aggregation]{\includegraphics[width=0.27\linewidth]{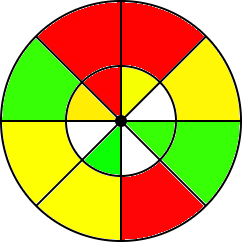}\label{fig:context_aggregated}}
    \caption{Context values creation with spatial transformation. Adapted from \cite{Yi2018}.}
    \label{fig:context}
\end{figure}

To encode spatial relationships we adopt the spatial transformation solution from \cite{Yi2018}, where the neighbourhood of a particular sensor is divided into $16$ areas as seen in Figure~\ref{fig:context_partitioning}. The mean values of these $16$ areas are used, together with the meteorological variables
as the context vector input for our networks. The wind direction is one-hot encoded to the directions used for the sensor measurement partitioning.

The neighbourhood representation also deals with mobile sensors implicitly. Since all sensors in an area are aggregated, there is no need for explicit handling of sensor placements, and thus do not need to separate between static and mobile sensors. This enables us to use both mobile and static sensors without changing pre-processing or model designs.

\section{Experiments}

\subsection{Experimental Setup}

\begin{table}
\renewcommand*{\arraystretch}{1.5}
    \centering
    \begin{tabular}{|c|c|c|c|}
    \hline
        Exp. & Train & Validation & Test \\ \hline \hline
        1 & 7 months & 1 month & 4 months \\ \hline
        2 & 3 months & \makecell{3 last weeks \\ from training} & 9 months \\ \hline
        3 & \makecell{8 months of \\ 5 drifted TS} & \makecell{Last 4 months \\ of 5 drifted TS} & \makecell{All 12 months \\ 6th drifted TS} \\ \hline
    \end{tabular}
    \caption{Dataset partition into training, validation and test sets for different experiments. \textbf{1}: standard, \textbf{2}: training on limited time, \textbf{3}: testing and training on different drift samples.}
    \label{tab:exp_setup}
\end{table}

Three experiments were set up: 1) standard; 2) limited training data; 3) drift generalization.
All models are optimized using the Adam optimizer with a learning rate of 0.001. The current learning rate is reduced to a factor of 0.2 when two epochs are completed with no loss decrease. The mean squared error (MSE) is used for optimization:
\[
MSE = (\frac{1}{n})\sum_{i=1}^{n}(y_{i} - x_{i})^{2}
\]

\paragraph{Standard}
A general performance test, in which models are trained on the first $7$ months, using the $8^{\text{th}}$ month for validation and early stopping. This provides a general performance metric on the synthetic data.
\paragraph{Limited training data}
A restrained experiment where the model trains on the first three months. By testing on the entire year, we get a metric describing the model’s ability to provide good calibration predictions for time-periods that are far away in time from the training data, and its ability to learn important features with limited data.
\paragraph{Drift generalization}
The model is trained on multiple simulated drift-values, and tests on another drift-value entirely, to provide insight into the model’s ability to generalize between drifts and be a more general model for this task.

\subsection{Results}

Table~\ref{tab:MSE_main} shows the MSE for the models in all experiments.
Conv2D is generally the best performing model with an exception of PM\textsubscript{10} drift prediction in the \textit{Limited training} experiment. Nonetheless, all new calibration models (Conv1D, Conv2D, LSTMwA) have errors in the same order of magnitude while consistently outperforming the baseline.

During experiments, the LSTMwA network had exploding gradients when training with the dataset containing multiple drift values. The reported MSE of that particular model-experiment combination is the best result obtained after $3$ tries.
Because of the repeated attempts on LSTMwA, the other models were also tested on the drifts experiment one second time, reporting the best score.


\newcolumntype{Y}{>{\centering\arraybackslash}X}
\begin{table}
    \centering
    \begin{tabularx}{\columnwidth}{|Y|Y|Y|Y|}
    \hline
        Model  & All & PM2.5 & PM10\\
        \hline
        \multicolumn{4}{|c|}{Standard} \\
        \hline
        Conv1D      & $0.0018$ & $0.0016$ & $0.0020$ \\
        Conv2D      & $\mathbf{0.0015}$ & $\mathbf{0.0011}$ & $\mathbf{0.0019}$ \\
        LSTMwA      & $0.0017$ & $0.0013$ & $0.0020$ \\
        Baseline & $0.0152$ & $0.0181$ & $0.0122$ \\
                \hline
        \multicolumn{4}{|c|}{Limited training} \\
        \hline
        Conv1D      & $0.0033$ & $0.0016$ & $0.0050$ \\
        Conv2D      & $\mathbf{0.0025}$ & $\mathbf{0.0013}$ & $0.0037$ \\
        LSTMwA      & $0.0038$ & $0.0042$ & $\mathbf{0.0034}$ \\
        Baseline & $0.0166$ & $0.0203$ & $0.0128$ \\
                \hline
        \multicolumn{4}{|c|}{Drift generalization} \\
        \hline
        Conv1D      & $0.0033$ & $0.0042$ & $0.0024$ \\
        Conv2D      & $\mathbf{0.0015}$ & $\mathbf{0.0017}$ & $\mathbf{0.0014}$ \\
        LSTMwA      & $0.0034$ & $0.0018$ & $0.0049$ \\
        Baseline & $0.0143$ & $0.0153$ & $0.0133$ \\
        \hline
    \end{tabularx}
    \caption{Performance on experiments for all models.}
    \label{tab:MSE_main}
\end{table}

\begin{figure}[tb]
    \centering
     \subfloat[\textbf{Baseline},Exp.1]{\includegraphics[width=0.44\linewidth]{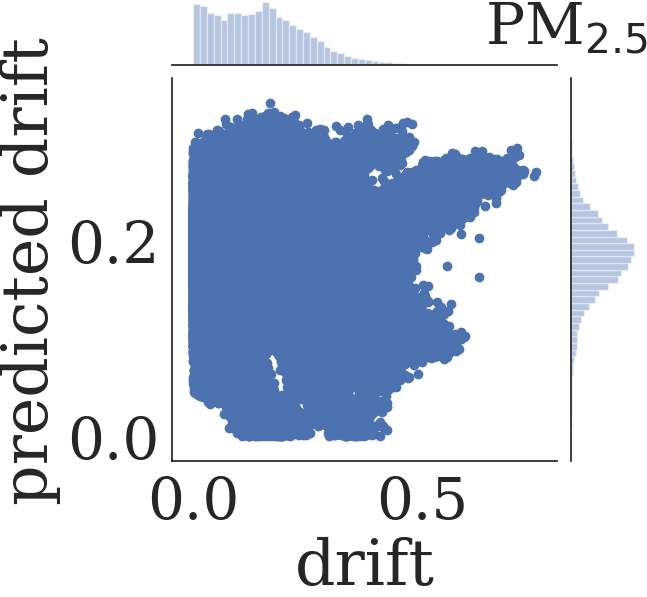}\label{fig:baseline_std_pm25}} \hfill
     \subfloat[\textbf{Conv2D},Exp.1]{\includegraphics[width=0.46\linewidth]{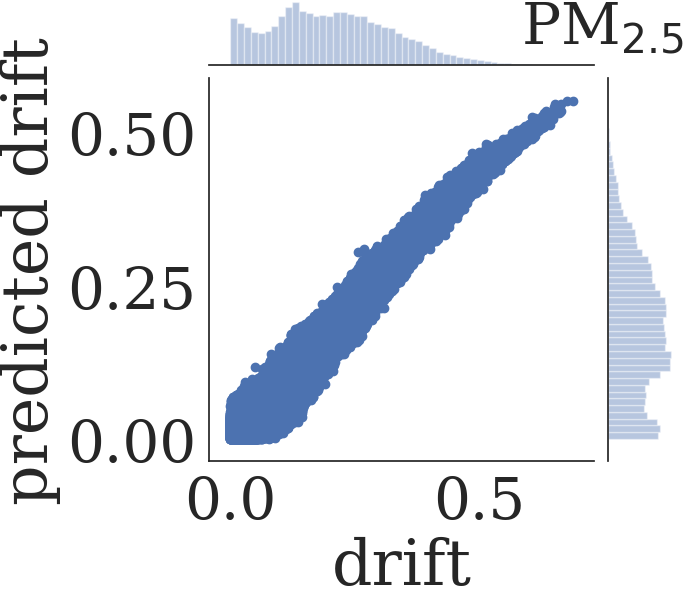}\label{fig:conv2d_std_pm25}} \\
     \subfloat[\textbf{Baseline},Exp.2]{\includegraphics[width=0.45\linewidth]{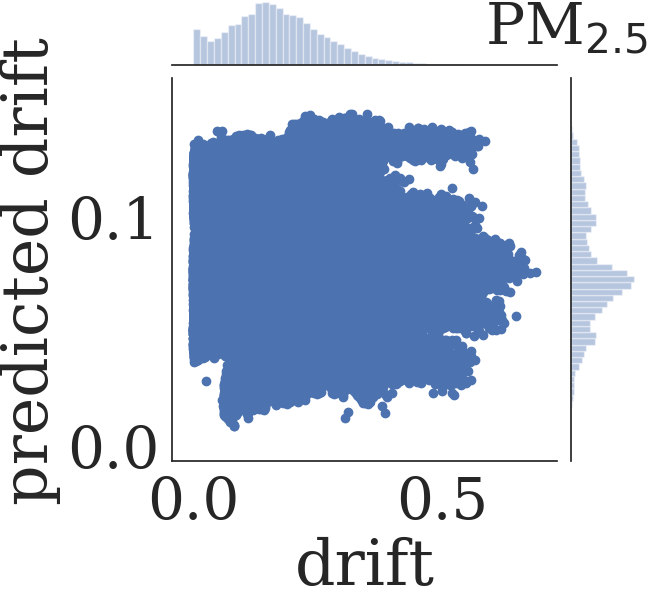}\label{fig:baseline_far_pm25}} \hfill
     \subfloat[\textbf{Conv2D},Exp.2]{\includegraphics[width=0.46\linewidth]{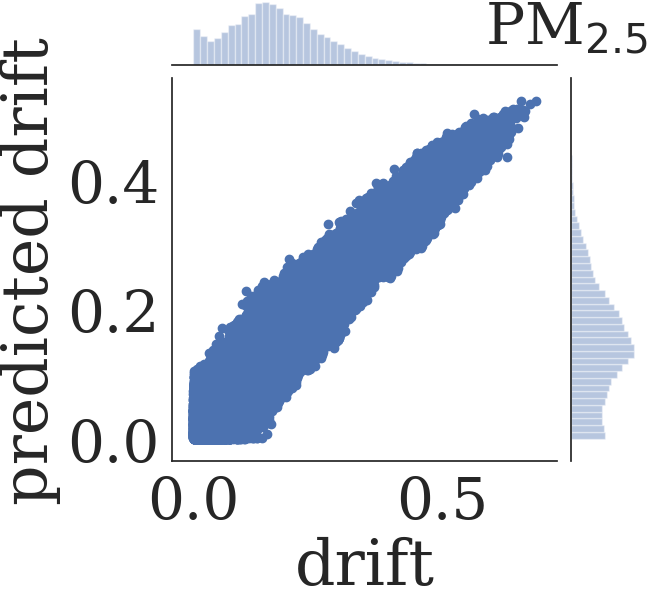}\label{fig:conv2d_far_pm25}} \\
     \subfloat[\textbf{Baseline},Exp.3]{\includegraphics[width=0.45\linewidth]{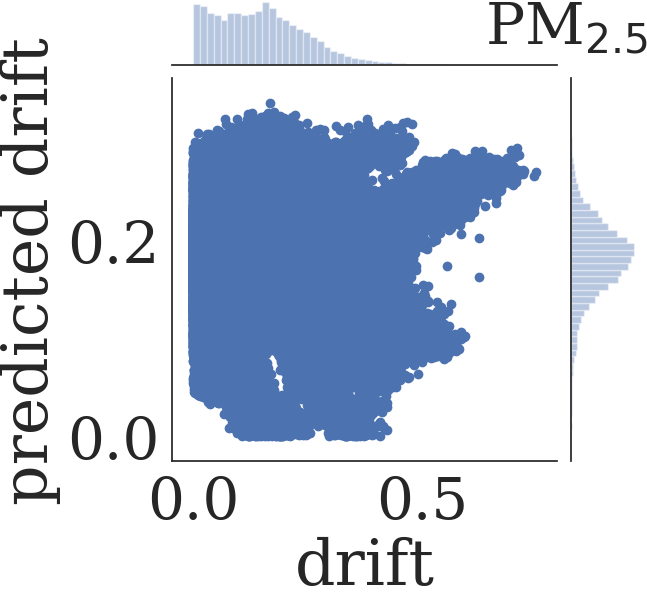}\label{fig:baseline_drift_pm25}} \hfill
     \subfloat[\textbf{Conv2D},Exp.3]{\includegraphics[width=0.48\linewidth]{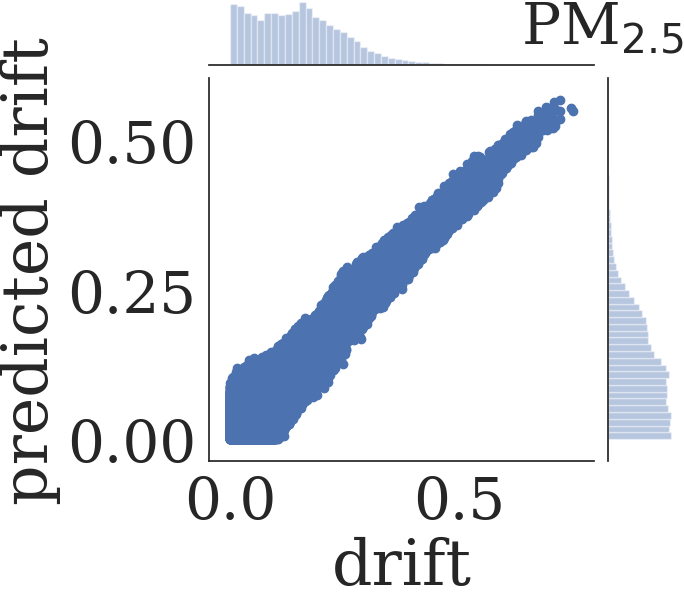}\label{fig:conv2d_drift_pm25}}
    \caption{Scatterplots comparing predicted vs true drift values. Predictions using baseline and Conv2D for PM2.5. \protect\subref{fig:baseline_std_pm25}, \protect\subref{fig:conv2d_std_pm25}: Standard; \protect\subref{fig:baseline_far_pm25}, \protect\subref{fig:conv2d_far_pm25}: Training on limited time; \protect\subref{fig:baseline_drift_pm25}, \protect\subref{fig:conv2d_drift_pm25}: Training on different drift samples.}
    \label{fig:scatter_plots_pm25}
\end{figure}

Figures~\ref{fig:scatter_plots_pm25}
shows how the predicted drift is correlated to the true drift for PM\textsubscript{2.5}, comparing our best performing model (Conv2D) against the baseline. Plots show almost no correlation between drift and predicted drift for the baseline model, while the opposite happens for Conv2D. We consistently observed this behavior also for PM\textsubscript{10}.


\def\figwidth{0.45\linewidth}

\subsection{Discussion}

Overall, the DL models introduced in this paper outperform the baseline and show good performance both with static and mobile sensors, with the Conv2D model consistently performing slightly better than others for all experiments.

The set of experiments show that our models adapt to different scenarios. The experiment with limited training data (Exp.2) shows that it is possible to calibrate sensors with a limited amount of training data available, allowing for drift calibration shortly after sensor deployment. Also, the experiment with drift generalization (Exp.3) shows that the calibration procedure can be transferred between different sensors, excluding the need for reference data for all the deployed sensors.

Our models show good performance both for PM\textsubscript{2.5} and PM\textsubscript{10}, showing that it is possible to calibrate more than one measurand per network. We also highlight that this behavior extends both for mobile and static sensor calibration, a novel contribution on the calibration of mobile sensors.

\section{Conclusions}


This paper explores using DL models for blind sensor calibration in air quality WSNs with static and mobile sensors.
Due to the effect of the COVID pandemic in the deployment of a local sensor network, and the lack of publicly available data for such sensor networks, we also propose a method to generate synthetic data for PM\textsubscript{2.5} and PM\textsubscript{10}, including sensor drift and the influence of meteorological variables in the drifted measurements.

Three new DL models are presented
and experiments show that all models outperform the baseline \cite{Wang2017} (adapted to deal with mobile sensors) on the synthetic data with a MSE of an order of magnitude less, with the $2$D convolutional model performing best by a small margin and  
with good performance both on the calibration of static and mobile sensors. This shows that DL for blind WSN calibration can be improved further, and possibly catch up to the more traditional methods without leveraging assumptions that may be inaccurate.

Our results show that remote automatic calibration of a network of drifting sensors is possible. In future work, we would like to consolidate these conclusions with testing on data from the deployed sensor network. There are a number of design and deployment decisions which might influence the performance of the sensors and the models. Some examples are the design of the sensor casing, the limited number of reference sensors to be used as ground truth or the location of the sensors to minimize the probability of external influence.




\bibliographystyle{IEEEtran}
\bibliography{main}

\end{document}